\documentclass[sigconf]{acmart}
\usepackage{tcolorbox}
\usepackage{enumitem}
\usepackage{xcolor}  
\usepackage{listings}
\usepackage{multirow}
\usepackage{tabularx}

\lstdefinestyle{stdout}{
    basicstyle=\ttfamily\small,
    backgroundcolor=\color{gray!10},
    frame=single,
    breaklines=true,
    columns=fullflexible,
    keepspaces=true
}

\AtBeginDocument{%
  }

\setcopyright{acmlicensed}
\copyrightyear{2026}
\acmYear{2026}
\acmDOI{XXXXXXX.XXXXXXX}
\acmConference[ASE'26]{IEEE/ACM International Conference on Automated Software Engineering}{October 12--16,
  2026}{Munich, Germany}
\acmISBN{978-1-4503-XXXX-X/2018/06}

\begin{document}

\title{An Exploration of Agentic Information Fusion for Test Maintenance Prediction}

\author{Jingxiong Liu}
\email{liujing@chalmers.se}
\orcid{0000-0002-5903-3179}
\affiliation{%
  \institution{Chalmers University of Technology and University of Gothenburg, Ericsson AB}
  \city{Gothenburg}
  \country{Sweden}
}

\author{Nasser Mohammadiha}
\orcid{0009-0000-4830-6250}
\affiliation{%
  \institution{Ericsson AB}
  \city{Gothenburg}
  \country{Sweden}
}

\author{Gregory Gay}
\email{greg@greggay.com}
\orcid{0000-0001-6794-9585}
\affiliation{%
  \institution{Chalmers University of Technology and University of Gothenburg}
  \city{Gothenburg}
  \country{Sweden}
}

\renewcommand{\shortauthors}{Liu et al.}

\begin{abstract}
Test maintenance is a critical, yet costly, activity---particularly as codebases rapidly evolve. To assist, we present MAST, a multi-agent framework that predicts which test cases require maintenance following changes to the production code. This identification task is necessary as a precondition to any subsequent maintenance activities, but remains challenging due to the complex relationships between production and test code. MAST advances the state-of-the-art by integrating multiple analyses---including static, lexical, and semantic analyses---through an intelligent fusion and post-check procedure and by focusing on a realistic use and evaluation setting---i.e., standardized input formats, repository-level analyses, and the ability to infer relations between test and production artifacts rather than assuming a pre-existing mapping. 

We evaluated MAST on 21 industrial Java repositories from Ericsson AB, considering situations where test maintenance both was and was not required in the ground truth. MAST yielded superior precision to a state-of-the-art baseline---resulting in a higher accuracy, F1, and F2 score---with only some loss in recall. Our ablation study demonstrates the value of each analysis in producing the final recommendations. MAST illustrates the potential of multi-agent systems that can fuse multiple information sources when performing software testing tasks.

\end{abstract}



\received{23 April 2026}
\received[accepted]{21 June 2026}

\maketitle

\section{Introduction}\label{sec:intro}
Software testing is a crucial, but expensive, stage in the quality assurance process~\cite{alegroth2016maintenance}. 
Although there is an initial cost associated with creating test cases, much of the cost of testing is imposed by the ongoing need for \textit{test maintenance}~\cite{sneed2004cost}, i.e., the adaptation of the test suite as the project evolves. In most cases, such activities takes place after changes are made to the source code that render the existing test suite obsolete---i.e., some tests may no longer be necessary, some may need adaptations to match the changed behavior of the component-under-test, or new tests may be needed~\cite{alegroth2016maintenance}. Test maintenance can also be conducted to improve the quality or efficiency of the test suite---for example, pruning redundant tests to reduce the execution time of the suite~\cite{sneed2004cost}. 

In current practice, test maintenance requires long-term commitment of significant developer effort~\cite{sneed2004cost,wang2021understanding}. Automation could assist in reducing the cost and improving the quality or reliability of the test maintenance process~\cite{imtiaz2019systematic,wang2021understanding}. In particular, in this study, we focus on how a \textit{multi-agent LLM system}---where a task is performed by a team of cooperating, semi-autonomous agents~\cite{He25:MultiAgent,liu2025exploringintegrationlargelanguage}---can assist in the test maintenance process. In a multi-agent LLM system, an overall task is split into sub-tasks that are performed by individual agents, where each agent pairs an LLM with instructions for the sub-task and access to tools, external data sources (e.g., via retrieval-augmented generation), and memory structures~\cite{He25:MultiAgent,liu2025exploringintegrationlargelanguage}. 

\begin{figure}[!h]
    \centering
    \includegraphics[width=\linewidth]{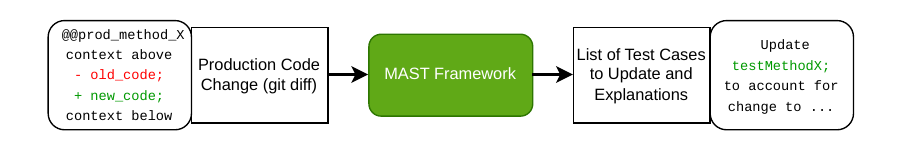}
    \caption{Input and output of the MAST framework.}
    \label{fig:motivational-example}
    \Description{A flow diagram consisting of the input (a production code change), connected to the MAST framework, then to the output (a list of test cases to update and explanations).}
\end{figure}

Regardless of the purpose of a test maintenance activity, one of the first steps in the process is to \textit{identify which existing test cases need to be modified or deleted}---a task we refer to as ``test localization''. In this study, we propose MAST (\textbf{M}ulti-\textbf{A}gent Multi-\textbf{S}ource \textbf{T}est Maintenance), a multi-agent framework that integrates multiple complementary analysis techniques to perform test localization following a change to the project source code. 

As shown in Figure~\ref{fig:motivational-example}, MAST takes as input a \texttt{git diff} showing a production code change, then outputs a list of test cases to modify or delete. For each test case, an explanation of why a change is needed is also provided. A tester could then go on to use this information to perform their planned maintenance activities. In the future, MAST could also be expanded to directly perform such actions on the identified test cases. In terms of scoping, we focus on code-based test cases at the unit, integration, and system-level. We have focused on Java-based projects in our evaluation, however, MAST is not conceptually tied to a particular language or testing framework. 

We have developed and evaluated MAST in collaboration with Ericsson AB, a major Swedish telecommunications company. MAST offers the following novel contributions over previous work (e.g., ~\cite{hu2023identify,chi2025reaccept}), including our own prior work~\cite{liu2025exploringintegrationlargelanguage}:

\smallskip\noindent\textbf{Multi-Source Analysis:} MAST yields predictions by fusing the results of three complementary analyses of the production code change and the test cases. First, \textit{static analysis}---specifically, call graph analysis---is used to identify which test cases invoke the changed production code. Second, \textit{lexical analysis} is used to tokenize and analyze the similarity of production and test code. Third, MAST performs a \textit{semantic analysis} where natural language summaries of the production code change and test cases are compared. 

The results of these analyses are then fused by an LLM agent, then the fused list is subjected to an additional \textit{post-check analysis}, where an agent compares the code of each potential candidate test to the production code change to filter false positives. Collectively, we hypothesize that these four analyses and the fusion procedure will yield more accurate results than previous studies---e.g., our prior work applied semantic analysis alone~\cite{liu2025exploringintegrationlargelanguage}. 
While LLM-based analysis fusion has been applied to other aspects of software development (e.g.,~\cite{GAO2025107803,10844085,khanghah2026mcerfadvancingmultimodalllm}), it has not been applied in test maintenance. 

\smallskip\noindent\textbf{Realistic Application Setting:} MAST has been designed and evaluated in an industrial context, with the aim of producing a tool that could be applied in practice. Unlike, for example,~\cite{hu2023identify}, MAST takes a \texttt{git diff} as input---a standard method of summarizing code changes---enabling direct invocation following a commit. While~\cite{hu2023identify} was limited to analyzing changes to a single production method, MAST can process repository-level changes. Previous approaches, e.g.,~\cite{chi2025reaccept} assumed a precise, pre-existing mapping of production and test artifacts, while MAST infers that relationship. Finally, we have examined the performance of MAST in both positive (i.e., in the ground truth, test cases were updated by developers following a production code change) and negative (i.e., production code changes did not lead to test maintenance) situations, ensuring that our framework can be used in situations where we do not already know whether maintenance is required. 

We evaluated MAST on 21 Java repositories from five teams at Ericsson, where we analyzed the performance of MAST in comparison to a state-of-the-art baseline derived from our previous research~\cite{liu2025exploringintegrationlargelanguage}. We also conducted an ablation study where we examined the impact of the individual analyses and the fusion on the overall performance of MAST. We observed that:
\begin{itemize}
    \item In cases where test maintenance was required, MAST achieves a much higher precision than the baseline, leading to improved accuracy, F1, and F2 scores. There is a trade-off between precision and recall, and the baseline achieves higher recall than MAST. 
    \item In cases where test maintenance was not required, MAST yields far fewer false positives than the baseline, resulting in a higher accuracy. 
    \item Each information source offers different maintenance suggestions. The fusion agent effectively merges the sources, and the post-check prunes many false positives from the merged list (albeit at some cost in recall). The superior results of MAST in precision, accuracy, and F1 score over the examined sub-workflows shows that each agent contributes to the overall effectiveness of the framework. However, if recall is highly prioritized over prevision, a tester may also wish to inspect the fusion results prior to the post-check. 
\end{itemize}

MAST demonstrates the potential of multi-agent LLM systems that can fuse multiple information sources when performing software testing tasks, such as test localization. Our findings advance our understanding of how multi-agent LLM systems can be applied within test maintenance and MAST represents a starting point for a future framework that can perform diverse maintenance tasks on the identified test cases. We make MAST available for researchers and practitioners to use and extend (see Section~\ref{sec:availability}).

\section{Background, Scoping, and Related Work}\label{sec:background}

\noindent\textbf{Software Testing:} Software testing is an activity performed to ensure that software delivers correct functionality and meets quality goals (e.g., performance)~\cite{Aniche22:Effective}. During testing, a \textit{test suite}---one or more \textit{test cases}---is executed against the system-under-test. Each test case performs input actions, then compares the resulting program behavior against expectations (called \textit{test oracles}). 

Testing can take place at multiple levels of granularity. While our work is not explicitly tied to a specific level, the projects used in the evaluation contain \textit{unit tests}---i.e., tests targeting an individual class---and \textit{integration tests}---i.e., tests targeting a set of collaborating units. These tests are written in the JUnit framework for Java\footnote{\url{https://junit.org/}}.

\smallskip\noindent\textbf{Test Maintenance:} Test maintenance is the process of updating an existing test suite~\cite{alegroth2016maintenance,liu2025exploringintegrationlargelanguage}. It often corresponds to a change in the production code, leading to addition of new tests and removal or adaptation of obsolete tests~\cite{alegroth2016maintenance,liu2025exploringintegrationlargelanguage}. Alternatively, maintenance may occur to improve efficiency (e.g., removal of redundant tests), due to changes in the technology stack or requirements, in response to discovery of a fault, or as a result of a deliberate decision to improve coverage or other quality indicators~\cite{liu2025exploringintegrationlargelanguage}. 

Test maintenance is understood to be important~\cite{kochhar2019practitioners} and can account for a significant proportion of testing effort over the lifespan of a project~\cite{alegroth2016maintenance,sneed2004cost}. However, in practice, few open-source projects implement tests following patterns that promote maintainability~\cite{gonzalez2017large} and the area has received comparatively little research attention~\cite{skoglund2004case}---particularly in industrial settings~\cite{imtiaz2019systematic}.

\begin{figure*}[!t]
    \centering
    \includegraphics[width=\linewidth]{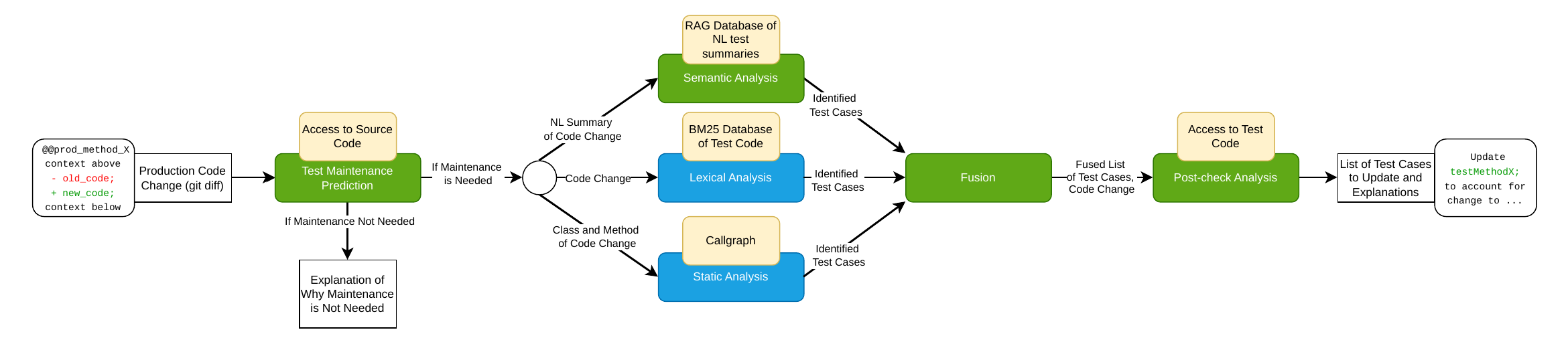}
    \caption{Overview of MAST. Boxes colored green represent agents, while the ones colored blue are scripted components.}
    \label{fig:system_overview}
    \Description{A flow diagram of the MAST framework. The production code change is passed to the Test Maintenance Prediction agent. If test maintenance is not needed, then an explanation is returned. Otherwise, the three analysis agents are invoked in parallel. Each returns a list of identified test cases. The Fusion agent fuses these into a combined list. The Post-Check Analysis then checks each test on this list against the production code change, outputting a final list of test maintenance recommendations.}
\end{figure*}

\smallskip\noindent\textbf{Large Language Models and Multi-Agent LLM Systems:} LLMs are a form of machine learning model designed to produce open-ended text-based responses to instructions (called \textit{prompts})~\cite{cao2023comprehensive}. LLMs are trained to infer the semantic meaning of a prompt and use that meaning to produce an appropriate response. LLMs are suited for language analysis and transformation tasks such as translation, summarization, and decision support~\cite{zhao2023survey}. Because of their ability to understand and output both natural language and code, LLMs are effective at software development and testing tasks~\cite{wang2024software}.

An \textit{LLM agent} is a software component coupling an LLM and a prompt with tool access---including access to external data sources---and memory capabilities~\cite{He25:MultiAgent,feldt2023towards}. These mechanisms allow LLM agents to reason, plan, and perceive or interact with an environment~\cite{feldt2023towards}. For example, an LLM agent may have access to a database of relevant context (e.g., source code) through Retrieval-Augmented Generation (RAG)~\cite{lewis2020retrieval}. This would allow it to reason about the code, make changes to it, and push those changes through access to the version control system. Multiple agents can work together---\textit{a multi-agent system}---each given different roles or sub-tasks~\cite{He25:MultiAgent}. The collaboration between agents (a \textit{workflow}) can be pre-planned by developers or agent-determined at execution time.

In terms of scoping, our multi-agent system consists only of pre-planned agent collaborations, with the workflow defined as a directed graph where each node represents an LLM agent or a \textit{scripted component}---functionality performed through traditional source code, rather than through invocation of an LLM. To simplify the presentation, we refer to all LLM-based nodes as ``agents'', even if that node does not include tool or external data access. 

\smallskip\noindent\textbf{Related Work:} Automated test maintenance has been under-studied in the research literature~\cite{hu2023identify,liu2025exploringintegrationlargelanguage}. However, test localization and obsolete test code repair have both been a focus of prior work. Traditional approaches are pattern-based, e.g., based on linking tests with production code purely through naming conventions~\cite{van2009establishing} or method signatures and return values~\cite{mirzaaghaei2010automatically,mirzaaghaei2011automatic,mirzaaghaei2014automatic}. 
Recently, there has been interest in applying LLMs to both localization and repair.

Liu et al.~\cite{liu2024fix}, Tu et al.~\cite{Tu2025:Repair}, and Xu et al.~\cite{Xu25:Update} have developed multi-agent or LLM-based frameworks for test repair. All three studies abstract the test localization problem, assuming that it is already known which tests require maintenance. Liu and Tu both employ static analyses to enhance their frameworks. Liu et al. use static analyses to infer context on the changed code class, how it is invoked, and the environment that the test and code are executed in~\cite{liu2024fix}. Tu et al. use a call graph to trace non-local dependencies~\cite{Tu2025:Repair}. Xu et al. use BM-25-based lexical analysis to retrieve similar code-test evolution pairs to use in formulating the test repair~\cite{Xu25:Update}. 


Of particular relevance are two LLM-related approaches that perform test localization. Hu et al. applied a transformer model to both localize and update tests~\cite{hu2023identify}. They do not separate these tasks into individual predictions, but perform a single repair task. If the final test is identical to the input, they assume that the test has been determined to still be relevant. Their approach is limited to production changes contained within a single method.

Recently, Chi et al. proposed a multi-agent framework for test localization~\cite{chi2025reaccept}. Their approach is conceptually similar to our ``post-check'' agent---comparing the test code to the production code change. However, their approach also employs a RAG-based memory mechanism containing natural language summaries of 
reasons why previous test updates took place. This is implemented in a similar manner to our semantic analysis, but focusing on different contextual information. A limitation of their approach is that it assumes that the set of all tests possibly affected by a production code change is already known, while our approach infers this relationship, and their evaluation dataset has a one-to-one mapping between production and test code changes. Neither approach considers commits where no test maintenance took place at all following a production change, only situations where particular individual test cases were not updated.

While LLM-based information fusion has been studied in other aspects of the development process---e.g., vulnerability detection~\cite{GAO2025107803}, test report clustering~\cite{10844085}, and documentation evaluation~\cite{khanghah2026mcerfadvancingmultimodalllm}---we are the first to apply this concept to test maintenance.



\section{MAST Framework}\label{sec:tool}
\UseRawInputEncoding

MAST (\textbf{M}ulti-\textbf{A}gent Multi-\textbf{S}ource \textbf{T}est Maintenance) is a multi-agent framework that integrates multiple analysis techniques to predict which test cases must be updated following a change to the production code. 
An overview of MAST is shown in Figure~\ref{fig:system_overview}, where the green boxes represent LLM agents and blue boxes represent scripted components. MAST executes the following workflow:
\begin{itemize}
    \item The \textbf{test maintenance prediction agent} (Section~\ref{sec:agent_maint}) assesses the production code change and predicts whether test maintenance is needed.
    \item Three analyses are executed in parallel---\textbf{semantic analysis}, \textbf{lexical analysis}, and \textbf{static analysis} (Section~\ref{sec:agent_analysis}). Each yields a list of candidate test cases.
    \item The \textbf{fusion agent} (Section~\ref{sec:agent_fusion}) merges these lists into a single, filtered, candidate list. 
    \item Then, the \textbf{post-check analysis agent} (Section~\ref{sec:agent_postcheck}) compares the code of each test on this list to the production code change, filtering those without clear relevance. 
\end{itemize}
This process yields a list of tests predicted to need maintenance, along with rationale for each inclusion.

\subsection{Input Format}

MAST uses the standard \texttt{git diff} format~\cite{mcquaid2014git} as input. This format shows code changes as well as surrounding unchanged lines for context. To offer additional context to MAST, we include nine code lines above and below each change. A single commit may consist of multiple \texttt{git diff} chunks. In such cases, we input one chunk at a time to MAST. An example \texttt{git diff} is shown at Listing~\ref{lst:gitdiff}. 

\begin{lstlisting}[style=stdout, basicstyle=\footnotesize, language=Java, caption={Example git diff chunk.}, label={lst:gitdiff}]
@@ -1446,20 +1446,20 @@ public final class AsciiString implements CharSequence, Comparable<CharSequence>;
             return false;
         }
-        for (int i = 0, j = 0; i &lt; a.length(); ++i, ++j) {
-            if (!equalsIgnoreCase(a.charAt(i),  b.charAt(j))) {
+        for (int i = 0; i &lt; a.length(); ++i) {
+            if (!equalsIgnoreCase(a.charAt(i),  b.charAt(i))) {
                 return false;
             }
\end{lstlisting}

\subsection{Test Maintenance Prediction Agent}\label{sec:agent_maint}

The test maintenance prediction agent uses an LLM to analyze the \texttt{git diff} chunk, extracting a natural language summary of the production code change and using the diff and summary to decide whether the change requires test maintenance. The agent will then offer, as output, its assessment along with the rationale and the code change summary.  An example is shown in Listing~\ref{lst:maintenance}.

If this agent decides that maintenance is not needed, then it will issue its output and conclude the execution of MAST. If it decides that maintenance is needed, the agent will also analyze the file where the diff chunk is situated and extract the method and class names from that file. It will then pass the rationale, change summary, and extracted method and class names to the analyses.

\begin{lstlisting}[style=stdout, basicstyle=\footnotesize, caption={Example maintenance prediction output.}, label={lst:maintenance}]
maintenance analysis is: {'needs_update': True, 
'maintenance_reason': 'The isAndroid() method implementation was changed from using a static variable IS_ANDROID to calling PlatformDependent0.isAndroid(). This change modifies the behavior and dependencies of the method, potentially affecting tests that mock or rely on the previous implementation.', 
'code_summary': 'File: PlatformDependent.java - Modified the isAndroid() method to delegate to PlatformDependent0.isAndroid() instead of returning a static IS_ANDROID variable'}
\end{lstlisting}

\subsection{Analysis Nodes}\label{sec:agent_analysis}

\subsubsection{Semantic Analysis Agent}

The intention of the semantic analysis is to analyze the underlying intent behind potentially relevant test cases and how that intent relates to the code change---i.e., what purpose does the test serve and what methods from the production code does it invoke? Semantic analysis has been proven to be powerful in solving software engineering problems (e.g.,~\cite{wang2021codet5}), especially when coupled with the inference capabilities of LLMs~\cite{zhang2024autocoderover,liu2025exploringintegrationlargelanguage}.

This agent uses an LLM equipped with RAG to retrieve natural language summaries of test cases from an embedding database (based on the \texttt{bge-m3} model~\cite{bge-m3}). These summaries are generated by an LLM for each test case in the suite, based on a commit-id. This database can be generated in advance or when MAST is invoked. An example of a test case and its summary can be seen in Listing~\ref{lst:raw code}. The agent retrieves the test summaries with the highest similarity to the natural language summary of the code change---i.e., those above a certain threshold using FAISS~\cite{johnson2019billion}. In this case, we retrieve test cases with similarity score above 0.50, with the values chosen through experimentation. We then perform a second filtering step, where we retain a subset with a similarity score $> mean + (0.25 * std. dev.)$, where the mean and standard deviation refer to the set being filtered.

\begin{lstlisting}[style=stdout, language=Java, showstringspaces=false,basicstyle=\footnotesize, caption={Example of a test case and its summary.}, label={lst:raw code}]
    @Test
    public void testGetBytesStringBuilder() {
        final StringBuilder b = new StringBuilder();
        for (int i = 0; i < 1 << 16; ++i) {
            b.append("eaà");
        }
        final String bString = b.toString();
        final Charset[] charsets = CharsetUtil.values();
        for (int i = 0; i < charsets.length; ++i) {
            final Charset charset = charsets[i];
            byte[] expected = bString.getBytes(charset);
            byte[] actual = new AsciiString(b, charset).toByteArray();
            assertArrayEquals("failure for " + charset, expected, actual);
        }
    }

Summary: Get bytes from StringBuilder with various charsets
Test name: testGetBytesStringBuilder()
Methods called: StringBuilder.append, StringBuilder.toString, CharsetUtil.values, String.getBytes, AsciiString constructor, AsciiString.toByteArray, assertArrayEquals
\end{lstlisting}


\subsubsection{Lexical Analysis}

Lexical analysis---an analysis based on parsing text into token vectors---is a classic method of similarity assessment~\cite{ghawi2019efficient}. To perform this analysis, we parse the raw source code of the test cases after a pre-processing stage where we convert all code to lower case, remove punctuation, and separate the keyword ``test'' from other words. 

We use the BM25 ranking algorithm~\cite{robertson2009probabilistic} to identify tests whose code is syntactically relevant to the code change. Due to its speed and simplicity, BM25 has also been applied for similar purposes in other LLM-based frameworks (e.g.,~\cite{wang2025rag,Xu25:Update,yang2025deep}). After parsing, the test code is stored in a BM25 index database. Then, all test cases that fall above a chosen similarity threshold will be retrieved from the index database. We retrieve up to five test cases with a similarity score above 50.00, values chosen through experimentation. 


\subsubsection{Static Analysis}

Static analysis encompasses a broad range of techniques that gather data from processing the production and test code~\cite{louridas2006static}. In particular, in this study, we use call graph analysis~\cite{1702621} to suggest test cases that are relevant to the production code change. Past research has shown that call graphs can reduce LLM hallucinations~\cite{gupta2025graphgroundedllmsleveraginggraphical,luo2025can,Tu2025:Repair}. 

\begin{lstlisting}[style=stdout, basicstyle=\footnotesize, caption={Example of the mapping of production methods to test cases from the call graph.}, label={lst:callgraph}]
"AsciiString.toByteArray": [
    "common/src/test/java/io/netty/util/AsciiStringCharacterTest.java::AsciiStringCharacterTest.testGetBytesStringBuilder",
    "common/src/test/java/io/netty/util/AsciiStringCharacterTest.java::AsciiStringCharacterTest.testGetBytesString",
    "common/src/test/java/io/netty/util/AsciiStringCharacterTest.java::AsciiStringCharacterTest.testGetBytesAsciiString"
  ]
\end{lstlisting}

This static analysis component uses the \texttt{tree-sitter} tool~\cite{treesitter} to parse the repository, and creates a call graph where each production method is mapped to all test cases that directly invoke that method. An example is shown in Listing~\ref{lst:callgraph}. Based on the methods extracted by the test maintenance prediction agent, this component extracts the potentially-impacted tests from the call graph. 

\subsection{Fusion Agent}\label{sec:agent_fusion}

The fusion agent employs an LLM to intelligently merge the test cases predicted by each of the three analysis nodes into a single list. As noted previously, LLM-based information fusion has shown potential as a flexible alternative to traditional heuristic-based fusion methods~\cite{GAO2025107803,10844085,khanghah2026mcerfadvancingmultimodalllm}. The prompt for this agent explains each information source, but does not suggest which sources may be more or less trustworthy to avoid biasing the model's predictions. The agent then returns a single list of test cases along with rationale to explain how it conducted the fusion process. An example of the rationale is shown in Listing~\ref{lst:fusion}. 

\begin{lstlisting}[style=stdout, basicstyle=\footnotesize, caption={Example of the fusion rationale.}, label={lst:fusion}]
The semantic analysis (Source 1) provides semantically relevant tests related to 
version parsing, security manager behavior, and appendable character sequence 
operations. From this list, 'testMajorVersion' and 
'testMajorVersionFromJavaSpecificationVersion' are further corroborated by 
Source 3 (lexical analysis), indicating lexical relevance as well. The append-related 
tests ('testAppendStringWithExpand', 'testSimpleAppendWithExpand', 
'testAppendString', 'testSimpleAppend') are also part of the semantic analysis 
results and fall under the same functional category, suggesting they exercise core 
logic likely impacted by changes in string or sequence handling. Source 2 (static 
analysis) does not provide additional tests beyond what's already included in 
Source 1, so it doesn't expand the result set. 'testIsZero' was only found via 
semantic analysis but lacks support from other sources and isn't clearly tied 
to the primary focus areas identified. Therefore, we select the six tests that 
show strong semantic and/or lexical relevance.
\end{lstlisting}

\subsection{Post-Check Analysis Agent}\label{sec:agent_postcheck}

The post-check analysis agent receives the fused list of test cases, then it extracts the code for each and compares the code to the production code change to assess the relevance of the suggested test. This agent generates a final, refined list of test cases. For each, it also provides rationale for its inclusion. An example of this rationale is shown in Listing~\ref{lst:postcheck}. 

\begin{lstlisting}[style=stdout, basicstyle=\footnotesize, caption={Example of post-check decision and reasoning.}, label={lst:postcheck}]
{'Update_or_Not': 'False', 
'post_check_reason': 'The test `testMajorVersionFromJavaSpecificationVersion` is designed to test the behavior of `PlatformDependent.majorVersionFromJavaSpecificationVersion()` under a custom `SecurityManager` that denies access to the `java.specification.version` system property. However, in the provided git diff, the methods `javaVersion0`, `majorVersionFromJavaSpecificationVersion`, and `majorVersion` have been removed from the `PlatformDependent` class. This means the method being tested no longer exists in the codebase. The test is no longer relevant and should not be retained without updates.'}
\end{lstlisting}

\subsection{Implementation Details}

MAST is implemented using LangGraph~\cite{langgraph2024}, a commonly-used open-source framework for creating multi-agent LLM systems. LangGraph represents the workflow of the system as a directed graph, where each node can be an agent or a traditional scripted component. 
The output of each agent---i.e., the current state---is stored in a common memory structure that can be accessed by all agents, like fetching information from a public information bus. The LangGraph architecture is highly flexible and scalable, which means that practitioners and researchers could add more agents to our existing framework---e.g., adding analyses or extending MAST to update the test code.

All agents that employ an LLM use the Qwen3-Coder-480B-A35B-Instruct-FP8 model~\cite{qwen3technicalreport} model, deployed on Ericsson's internal computing infrastructure. Ericsson must approve models for internal use, and this model was the best available at the time that we implemented MAST. This model is generally effective at coding tasks (e.g., achieving a 44\% resolved rate on SWE-rebench\footnote{\url{https://swe-rebench.com/}}). 

\section{Methods}\label{sec:methods}
\UseRawInputEncoding

\begin{table*}[!t]
\centering
\caption{Overview of projects used to evaluate MAST, including the name, number of positive and negative commits, average test suite size across all commits, and a description. Repository names are anonymized to preserve confidentiality.}
\label{tab:repos}
\scriptsize
\begin{tabular}{lcccl}
\toprule
\textbf{Project} & \textbf{Positive Commits} & \textbf{Negative Commits} & \textbf{Avg. Suite Size} & \textbf{Description} \\
\midrule
*-dataexport-test-stability& 2 & 0 & 2.00 & Support for stability testing of the data export tool. \\
*-extractor & 9 & 6 & 6.10 & Data extraction utility.  \\ 
*-dataexport-test-e2e & 9 & 2 & 8.40 & Support for end-to-end testing of the data export tool.\\
*-extractor-2 & 13 & 4 & 10.00 & Data extraction utility.   \\
*-service-authentication & 8 & 5 & 12.70 & Authentication service.\\
*-parser & 13 & 10 & 13.20 & Data parsing tool.  \\
*-datastream-*-scanner & 14 & 10 & 13.40 & Datastream scanner that scans and moves files and sends events.  \\
*-lib-common & 6 & 10 & 15.30 & Library of common utility functions. \\
*-datastream-dump-*-extractor & 3 & 2 & 16.00 & Extracts data dumps. \\
*-datastream-dump-cleaner & 16 & 10 & 19.30 & File system cleanup utility.\\
*-datastream-archive & 16 & 10 & 40.40 & Tool that archives data streams. \\
*-service-dataexport-download & 8 & 5 & 42.60 & Download service used to export data.  \\
archiver & 54 & 10 & 49.80 &  Archiver service. \\
*-datastream-dump-scanner & 78 & 10 & 49.90 & File system scanner that detects new files and publishes information about them.  \\
*-service-dataexport-executor & 31 & 10 & 67.90 & Provides an API for a UI to manage data export execution.\\
*-decrypter & 8 & 6 & 87.30 &  Decrypting utility. \\
*-datastream-common & 7 & 10 & 93.50 & A common library for all datastream-based projects.\\
common & 5 & 2 & 97.20 & Common library for a data ingestion service.  \\
*-datastream-dump-common & 72 & 10 & 103.60 & A common libray for all the datastream dump  projects.\\
*-datastream-dump-archiver & 123 & 10 & 114.00 & Archiver component that reacts to  messages about new files available on the file system.\\
*-service-dataexport-handler & 55 & 10 & 118.00 & Data export service. \\
\bottomrule
\end{tabular}
\end{table*}

We have performed an evaluation of MAST on datasets curated from 21 Java repositories at Ericsson AB containing unit and integration test cases. Specifically, in this evaluation, we address the following research questions:
\begin{tcolorbox}[
    title=Research Questions,
    fonttitle=\bfseries,
    colback=gray!5,
    colframe=black!60,
    arc=2pt,
    boxrule=0.5pt,
    left=8pt, right=8pt, top=6pt, bottom=6pt,
    width=\linewidth
]
\begin{enumerate}[label=\textbf{RQ\arabic*:}, leftmargin=*, nosep, itemsep=4pt]
    \item What is the performance of MAST on \textit{positive} cases (i.e., where test maintenance was required in the ground truth), compared to the baseline approach?
    \item What is the performance of MAST on \textit{negative} cases (i.e., where test maintenance was not required), compared to the baseline approach?
    \item How do the individual components of MAST contribute to the performance of the overall framework? 
\end{enumerate}
\end{tcolorbox}

We split our analysis of each dataset into \textit{positive} cases (\textbf{RQ1}), where a production code change is followed by subsequent test maintenance, and \textit{negative} cases (\textbf{RQ2}), where a production code change did not require test maintenance. This is because a developer may not know in advance whether maintenance will be required, and MAST should yield acceptable performance in both situations. In \textbf{RQ3}, we conducted an ablation study where we compare the performance of different sub-workflows (individual analyses, fusion, post-check) within MAST to understand how they contribute to the overall performance of the framework. 

\subsection{Datasets}

We have extracted datasets for this evaluation from 21 Java repositories from five teams at Ericsson AB. A description and metadata for each repository is shown in Table~\ref{tab:repos}. The repositories are sorted by the average size of the test suite across the commits. 

To obtain the dataset for each repository, we extracted the commit history until Feb. 24th, 2026 of the main branch. To identify positive commits (\textbf{RQ1,3}), we filtered each commit to identify those where test cases were updated. Our aim was to identify cases where test updates are related to a meaningful change to the production code. Therefore, to ensure the quality of the identified commits, we then manually removed all commits where the test changes are unrelated to the source code change or are only superficial changes. Specifically, we removed commits where test changes were only renaming of test cases, changes due to linting requirements, changes imposed by a change in the test support code (rather than in the production code), or where the directory structure of the project was changed. After this filtering, we obtained 430 valid commits showing code-test co-evolution across the 21 repositories.

To obtain negative examples to address \textbf{RQ2}, we also selected the latest 10 commits from each repository where there was a production code change, but no changes to the test methods. In some cases, there were fewer than 10 commits without test updates. In such cases, we included all relevant commits. 

For each commit, we extracted the code of all test methods that were updated by the developers. We extracted the production code changes in \texttt{git diff} format, separating each change into different chunks. Each chunk includes, as additional context, the nine lines above and below the code changes. 

\subsection{Baseline Approach}

To compare MAST to the existing state-of-the-art for LLM-based test localization, we utilize a baseline representing our prior research~\cite{liu2025exploringintegrationlargelanguage}. In our previous study, we proposed a multi-agent LLM system for test localization based on semantic analysis alone. We have derived an equivalent implementation to the framework from our previous study by extracting a portion of the MAST pipeline consisting of the Maintenance Prediction agent and the Semantic Analysis agent from Figure~\ref{fig:system_overview}. 

\subsection{Data Collection}

To analyze performance, we executed MAST for each commit.
\begin{enumerate}
    \item The information stored in the semantic RAG database, BM25 index database, and call graph were updated to reflect the test suite prior to any test changes made in that commit, associated with a unique commit-id. In case of a re-execution, the contents of these databases or call graphs do not need to be regenerated.
    \item Each individual code change (\texttt{git diff} chunk) for that commit was input into MAST. 
    \item Ground truth is known for a commit, not for each chunk. To assess performance, the identified test cases for all chunks in that commit were aggregated, then compared to the ground truth for the actual commit.
\end{enumerate}

We executed all agents with a temperature setting of 0.00 to help ensure stability and result reproducibility. However, we performed the evaluation three times, as there may still be some variance in results. We present the mean and standard deviation across the three executions. For each commit, we define:
\begin{itemize}
    \item A \textbf{true positive (TP)} is a test predicted as \textit{requiring} maintenance that \textit{actually} required maintenance.
    \item A \textbf{true negative (TN)} is a test predicted as \textit{not requiring} maintenance that \textit{actually} did not require maintenance.
    \item A \textbf{false positive (FP)} is a test predicted as \textit{requiring} maintenance that \textit{did not} require maintenance.
    \item A \textbf{false negative (FN)} is a test predicted as \textit{not requiring} maintenance that \textit{actually}  required maintenance.
\end{itemize}

We collect results for the full MAST pipeline. In addition, for the baseline and ablation study, we also extract performance for the following sub-workflows within MAST:
\begin{enumerate}
    \item \textbf{Baseline (Semantic Only):} Maintenance Prediction, Semantic Analysis
    \item \textbf{Lexical Only:} Maintenance Prediction, Lexical Analysis
    \item \textbf{Static Only:} Maintenance Prediction, Static Analysis
    \item \textbf{Without Post-Check:} Maintenance Prediction, Semantic Analysis, Lexical Analysis, Static Analysis, Fusion
\end{enumerate}

Workflows 1--3 represent the three information sources alone, with the first representing the framework in our prior work~\cite{liu2025exploringintegrationlargelanguage}. Workflow 4, then, represents the framework without the final post-check analysis, with the three information sources and the fusion agent. Isolating these four workflows allows us to understand how accurate each individual analysis is and how each contributes to the overall performance of MAST. 

\subsection{Data Analysis}

We collect the number of true and false positives and negatives for MAST and each workflow for each commit. We then aggregate the results to calculate the following metrics:
\begin{itemize}
    \item Accuracy = $\frac{TP+TN}{(TP+FP+TN+FN)}$
    \item Recall = $\frac{TP}{(TP+FN)}$
    \item Precision = $\frac{TP}{(TP+FP)}$
    \item F1 score = $2 * \frac{(precision * recall)}{(precision + recall)}$
    \item F2 score = $(1 + 2^2)*  \frac{(precision * recall)}{(2^2*precision + recall)}$
\end{itemize}
Depending on the research question, we either aggregate TP, FP, TN, and FN across commits for each individual project---i.e., calculate these metrics per-project---or we aggregate TP, FP, TN, and FN across all commits for all projects---i.e., calculate metrics across the full set of commits. In the latter case, we aggregate to account for the imbalance in the number of commits per-project. 

We use these measurements to provide multiple viewpoints on performance. In particular, as some of the commits contain a large number of TN (i.e., a large test suite where few tests are updated), the accuracy may be artificially inflated. Therefore, the F1 and F2 scores are useful because they focus on TP, FP, and FN. While the F1 score is commonly used to assess machine learning tasks, we also include the F2 score---which places a heavier weight on recall than precision---because omitting tests that should have been included in a prediction is likely more detrimental than including some FPs.
We use these metrics to address the research questions as follows:
\begin{itemize}
    \item To address \textbf{RQ1}, we consider the set of positive commits (i.e., commits where test maintenance was required). We report all metrics for each project individually, as well as aggregated across all projects, for MAST and the baseline.  
    \item To address \textbf{RQ2}, we focus on negative commits (i.e., cases where no test maintenance was performed). We only report accuracy, as there can be no TP or FN, only TN and FP. Therefore, the other metrics cannot be calculated. We report accuracy across all projects for MAST and the baseline. We omit results for individual projects due to the relatively small number of negative commits extracted from each individual project. 
    \item To address \textbf{RQ3}, we report all five metrics for positive commits and accuracy alone for negative commits--- aggregated across all datasets---for MAST and all four workflows (including the baseline) defined in the previous section. We omit results for individual projects due to space constraints, but they are available in our supplemental data (Section~\ref{sec:availability}).
\end{itemize}

\begin{table*}[!t]
\centering
\caption{Comparison of the performance of MAST and the baseline for positive commits (maintenance required) on each project and aggregated across all projects. The best approach is in \textbf{bold} for each metric and repository.}
\label{tab:results_rq1}
\scriptsize
\begin{tabular}{llccccc}
\toprule
\textbf{Repository} & \textbf{Approach} & \textbf{Precision} & \textbf{Recall} & \textbf{Accuracy} & \textbf{F1 Score} & \textbf{F2 Score}\\
\midrule
\multirow{2}{*}{*-dataexport-test-stability} & MAST & \textbf{1.000 $\pm$ 0.000} & \textbf{1.000 $\pm$ 0.000} & \textbf{1.000 $\pm$ 0.000} & \textbf{1.000 $\pm$ 0.000} & \textbf{1.000 $\pm$ 0.000} \\
 & Baseline & 0.667 $\pm$ 0.289 & \textbf{1.000 $\pm$ 0.000} & 0.667 $\pm$ 0.289 & 0.778 $\pm$ 0.192 & 0.889 $\pm$ 0.096 \\
\midrule
\multirow{2}{*}{*-extractor} & MAST & \textbf{0.697 $\pm$ 0.022} & \textbf{0.929 $\pm$ 0.000} & \textbf{0.864 $\pm$ 0.012} & \textbf{0.796 $\pm$ 0.014} & \textbf{0.871 $\pm$ 0.007} \\
 & Baseline & 0.460 $\pm$ 0.024 & \textbf{0.929 $\pm$ 0.000} & 0.667 $\pm$ 0.031 & 0.615 $\pm$ 0.022 & 0.771 $\pm$ 0.014 \\
\midrule
\multirow{2}{*}{*-dataexport-test-e2e} & MAST & \textbf{0.775 $\pm$ 0.015} & 0.885 $\pm$ 0.000 & \textbf{0.836 $\pm$ 0.010} & \textbf{0.826 $\pm$ 0.009} & \textbf{0.860 $\pm$ 0.004} \\
 & Baseline & 0.500 $\pm$ 0.000 & \textbf{0.923 $\pm$ 0.000} & 0.559 $\pm$ 0.000 & 0.649 $\pm$ 0.000 & 0.789 $\pm$ 0.000 \\
\midrule
\multirow{2}{*}{*-extractor-2} & MAST & \textbf{0.631 $\pm$ 0.024} & \textbf{0.750 $\pm$ 0.054} & \textbf{0.817 $\pm$ 0.017} & \textbf{0.685 $\pm$ 0.035} & \textbf{0.723 $\pm$ 0.045} \\
 & Baseline & 0.444 $\pm$ 0.016 & \textbf{0.750 $\pm$ 0.054} & 0.683 $\pm$ 0.014 & 0.558 $\pm$ 0.025 & 0.659 $\pm$ 0.038 \\
\midrule
\multirow{2}{*}{*-service-authentication} & MAST & \textbf{0.611 $\pm$ 0.048} & 0.517 $\pm$ 0.029 & \textbf{0.785 $\pm$ 0.015} & \textbf{0.559 $\pm$ 0.016} & 0.532 $\pm$ 0.021 \\
 & Baseline & 0.436 $\pm$ 0.041 & \textbf{0.750 $\pm$ 0.050} & 0.675 $\pm$ 0.042 & 0.550 $\pm$ 0.023 & \textbf{0.654 $\pm$ 0.021} \\
\midrule
\multirow{2}{*}{*-parser} & MAST & \textbf{0.692 $\pm$ 0.031} & 0.923 $\pm$ 0.038 & \textbf{0.920 $\pm$ 0.013} & \textbf{0.791 $\pm$ 0.034} & \textbf{0.865 $\pm$ 0.036} \\
 & Baseline & 0.427 $\pm$ 0.013 & \textbf{0.974 $\pm$ 0.022} & 0.781 $\pm$ 0.013 & 0.594 $\pm$ 0.009 & 0.775 $\pm$ 0.004 \\
\midrule
\multirow{2}{*}{*-datastream-*-scanner} & MAST & \textbf{0.777 $\pm$ 0.033} & 0.658 $\pm$ 0.031 & \textbf{0.878 $\pm$ 0.013} & \textbf{0.712 $\pm$ 0.031} & 0.678 $\pm$ 0.031 \\
 & Baseline & 0.420 $\pm$ 0.020 & \textbf{0.856 $\pm$ 0.056} & 0.696 $\pm$ 0.019 & 0.564 $\pm$ 0.029 & \textbf{0.709 $\pm$ 0.041} \\
\midrule
\multirow{2}{*}{*-lib-common} & MAST & \textbf{0.846 $\pm$ 0.037} & 0.596 $\pm$ 0.067 & \textbf{0.710 $\pm$ 0.038} & 0.698 $\pm$ 0.049 & 0.633 $\pm$ 0.061 \\
 & Baseline & 0.687 $\pm$ 0.000 & \textbf{0.885 $\pm$ 0.000} & 0.707 $\pm$ 0.000 & \textbf{0.773 $\pm$ 0.000} & \textbf{0.836 $\pm$ 0.000} \\
\midrule
\multirow{2}{*}{*-datastream-dump-*-extractor} & MAST & \textbf{0.733 $\pm$ 0.208} & 0.615 $\pm$ 0.077 & \textbf{0.826 $\pm$ 0.087} & 0.666 $\pm$ 0.135 & 0.634 $\pm$ 0.098 \\
 & Baseline & 0.568 $\pm$ 0.019 & \textbf{0.872 $\pm$ 0.044} & 0.785 $\pm$ 0.012 & \textbf{0.687 $\pm$ 0.001} & \textbf{0.787 $\pm$ 0.021} \\
\midrule
\multirow{2}{*}{*-datastream-dump-cleaner} & MAST & \textbf{0.754 $\pm$ 0.035} & 0.667 $\pm$ 0.031 & \textbf{0.873 $\pm$ 0.014} & \textbf{0.708 $\pm$ 0.033} & 0.682 $\pm$ 0.032 \\
 & Baseline & 0.425 $\pm$ 0.004 & \textbf{0.884 $\pm$ 0.012} & 0.697 $\pm$ 0.005 & 0.574 $\pm$ 0.004 & \textbf{0.727 $\pm$ 0.006} \\
\midrule
\multirow{2}{*}{*-datastream-archive} & MAST & \textbf{0.763 $\pm$ 0.042} & 0.637 $\pm$ 0.027 & \textbf{0.939 $\pm$ 0.007} & \textbf{0.694 $\pm$ 0.032} & 0.659 $\pm$ 0.028 \\
 & Baseline & 0.391 $\pm$ 0.017 & \textbf{0.895 $\pm$ 0.030} & 0.837 $\pm$ 0.009 & 0.544 $\pm$ 0.021 & \textbf{0.711 $\pm$ 0.024} \\
\midrule
\multirow{2}{*}{*-service-dataexport-download} & MAST & \textbf{1.000 $\pm$ 0.000} & 0.389 $\pm$ 0.020 & 0.776 $\pm$ 0.007 & 0.560 $\pm$ 0.020 & 0.443 $\pm$ 0.020 \\
 & Baseline & 0.809 $\pm$ 0.020 & \textbf{0.684 $\pm$ 0.037} & \textbf{0.825 $\pm$ 0.010} & \textbf{0.740 $\pm$ 0.020} & \textbf{0.705 $\pm$ 0.030} \\
\midrule
\multirow{2}{*}{archiver} & MAST & \textbf{0.650 $\pm$ 0.009} & 0.635 $\pm$ 0.019 & \textbf{0.868 $\pm$ 0.004} & \textbf{0.642 $\pm$ 0.014} & \textbf{0.638 $\pm$ 0.017} \\
 & Baseline & 0.418 $\pm$ 0.010 & \textbf{0.723 $\pm$ 0.006} & 0.761 $\pm$ 0.007 & 0.530 $\pm$ 0.009 & 0.631 $\pm$ 0.008 \\
\midrule
\multirow{2}{*}{*-datastream-dump-scanner} & MAST & \textbf{0.612 $\pm$ 0.021} & 0.641 $\pm$ 0.027 & \textbf{0.915 $\pm$ 0.005} & \textbf{0.626 $\pm$ 0.024} & \textbf{0.635 $\pm$ 0.026} \\
 & Baseline & 0.316 $\pm$ 0.007 & \textbf{0.665 $\pm$ 0.011} & 0.803 $\pm$ 0.004 & 0.429 $\pm$ 0.008 & 0.545 $\pm$ 0.010 \\
\midrule
\multirow{2}{*}{*-service-dataexport-executor} & MAST & \textbf{0.449 $\pm$ 0.020} & \textbf{0.721 $\pm$ 0.063} & \textbf{0.911 $\pm$ 0.005} & \textbf{0.554 $\pm$ 0.034} & \textbf{0.643 $\pm$ 0.048} \\
 & Baseline & 0.208 $\pm$ 0.008 & 0.716 $\pm$ 0.017 & 0.770 $\pm$ 0.008 & 0.323 $\pm$ 0.011 & 0.481 $\pm$ 0.014 \\
\midrule
\multirow{2}{*}{*-decrypter} & MAST & \textbf{0.656 $\pm$ 0.028} & 0.617 $\pm$ 0.041 & \textbf{0.914 $\pm$ 0.007} & \textbf{0.635 $\pm$ 0.031} & \textbf{0.624 $\pm$ 0.037} \\
 & Baseline & 0.356 $\pm$ 0.011 & \textbf{0.631 $\pm$ 0.008} & 0.817 $\pm$ 0.006 & 0.455 $\pm$ 0.011 & 0.546 $\pm$ 0.010 \\
\midrule
\multirow{2}{*}{*-datastream-common} & MAST & \textbf{0.865 $\pm$ 0.045} & 0.889 $\pm$ 0.048 & \textbf{0.995 $\pm$ 0.002} & \textbf{0.877 $\pm$ 0.042} & \textbf{0.884 $\pm$ 0.044} \\
 & Baseline & 0.279 $\pm$ 0.006 & \textbf{1.000 $\pm$ 0.000} & 0.945 $\pm$ 0.002 & 0.436 $\pm$ 0.008 & 0.659 $\pm$ 0.007 \\
\midrule
\multirow{2}{*}{common} & MAST & \textbf{0.917 $\pm$ 0.061} & 0.464 $\pm$ 0.005 & \textbf{0.841 $\pm$ 0.009} & \textbf{0.616 $\pm$ 0.012} & 0.515 $\pm$ 0.004 \\
 & Baseline & 0.545 $\pm$ 0.011 & \textbf{0.645 $\pm$ 0.009} & 0.754 $\pm$ 0.006 & 0.591 $\pm$ 0.008 & \textbf{0.622 $\pm$ 0.008} \\
\midrule
\multirow{2}{*}{*-datastream-dump-common} & MAST & \textbf{0.711 $\pm$ 0.016} & 0.603 $\pm$ 0.004 & \textbf{0.971 $\pm$ 0.001} & \textbf{0.653 $\pm$ 0.009} & 0.622 $\pm$ 0.006 \\
 & Baseline & 0.394 $\pm$ 0.006 & \textbf{0.732 $\pm$ 0.014} & 0.937 $\pm$ 0.001 & 0.513 $\pm$ 0.008 & \textbf{0.625 $\pm$ 0.011} \\
\midrule
\multirow{2}{*}{*-datastream-dump-archiver} & MAST & \textbf{0.521 $\pm$ 0.003} & 0.623 $\pm$ 0.020 & \textbf{0.934 $\pm$ 0.001} & \textbf{0.567 $\pm$ 0.009} & \textbf{0.599 $\pm$ 0.015} \\
 & Baseline & 0.320 $\pm$ 0.007 & \textbf{0.678 $\pm$ 0.008} & 0.878 $\pm$ 0.002 & 0.435 $\pm$ 0.008 & 0.554 $\pm$ 0.009 \\
\midrule
\multirow{2}{*}{*-service-dataexport-handler} & MAST & \textbf{0.729 $\pm$ 0.011} & 0.522 $\pm$ 0.022 & \textbf{0.950 $\pm$ 0.002} & \textbf{0.608 $\pm$ 0.017} & 0.554 $\pm$ 0.021 \\
 & Baseline & 0.404 $\pm$ 0.008 & \textbf{0.646 $\pm$ 0.025} & 0.903 $\pm$ 0.002 & 0.497 $\pm$ 0.013 & \textbf{0.577 $\pm$ 0.019} \\
\midrule
\multirow{2}{*}{\textbf{Overall}} & MAST & \textbf{0.621 $\pm$ 0.003} & 0.613 $\pm$ 0.014 & \textbf{0.932 $\pm$ 0.001} & \textbf{0.617 $\pm$ 0.009} & \textbf{0.615 $\pm$ 0.012} \\
 & Baseline & 0.367 $\pm$ 0.001 & \textbf{0.708 $\pm$ 0.005} & 0.865 $\pm$ 0.000 & 0.483 $\pm$ 0.002 & 0.597 $\pm$ 0.003 \\
\bottomrule
\end{tabular}
\end{table*}

\section{Results}\label{sec:results}

\begin{table*}[!t]
\centering
\caption{Ablation study on positive and negative commits, aggregated across all projects.}
\label{tab:results_rq2_3}
\scriptsize
\begin{tabular}{lccccc|cc}
\toprule
& \multicolumn{5}{c|}{\textbf{Positive Commits}} & \multicolumn{2}{c}{\textbf{Negative Commits}} \\
\textbf{Workflow} & \textbf{Precision} & \textbf{Recall} & \textbf{Accuracy} & \textbf{F1 Score} & \textbf{F2 Score} & \textbf{False Positives} & \textbf{Accuracy} \\
\midrule
Baseline (Semantic-Only) & 0.367 $\pm$ 0.001 & 0.708 $\pm$ 0.005 & 0.865 $\pm$ 0.000 & 0.483 $\pm$ 0.002 & 0.597 $\pm$ 0.003 & 688.700 $\pm$ 19.500 & 0.931 $\pm$ 0.004 \\
Lexical-Only & 0.395 $\pm$ 0.002 & 0.458 $\pm$ 0.003 & 0.889 $\pm$ 0.000 & 0.424 $\pm$ 0.002 & 0.444 $\pm$ 0.002 & 200.300 $\pm$ 25.000 & 0.980 $\pm$ 0.003\\
Static Analysis-Only & 0.502 $\pm$ 0.003 & 0.590 $\pm$ 0.005 & 0.911 $\pm$ 0.001 & 0.542 $\pm$ 0.003 & 0.570 $\pm$ 0.004 & 188.700 $\pm$ 4.500 & 0.981 $\pm$ 0.000 \\
Fusion Without Post-Check & 0.355 $\pm$ 0.004 & \textbf{0.778 $\pm$ 0.018} & 0.854 $\pm$ 0.001 & 0.487 $\pm$ 0.007 & \textbf{0.628 $\pm$ 0.012} & 769.000 $\pm$ 8.200 & 0.923 $\pm$ 0.003\\
MAST (All Analyses) & \textbf{0.621} $\pm$ 0.003 & 0.613 $\pm$ 0.014 & \textbf{0.932 $\pm$ 0.001} & \textbf{0.617 $\pm$ 0.009} & 0.615 $\pm$ 0.012 & \textbf{113.300 $\pm$ 36.500} & \textbf{0.989 $\pm$ 0.002}\\
\bottomrule
\end{tabular}
\end{table*}

\subsection{Comparison of MAST and Baseline in Positive Cases (RQ1)}

Table~\ref{tab:results_rq1} lists the mean precision, recall, accuracy, F1 score, and F2 score of MAST and the baseline for the dataset extracted from each project, as well as aggregated across all projects. As we executed each approach three times to assess the variance of each approach, we also list the standard deviation. To answer this question, we focus on the positive commits---i.e., commits where test maintenance was necessary. For each metric, we list the best approach in \textbf{bold}. 

Overall, from Table~\ref{tab:results_rq1}, we see that MAST achieves a substantially higher precision, accuracy, F1, and F2 score than the baseline. The baseline represents only a single analysis performed in MAST, the semantic analysis, and the results suggest benefits from the additional analyses of information sources as well as the fusion procedure and the post-check analysis. The improvement in performance comes with a trade-off in recall, and the baseline achieves approximately 16\% higher recall than MAST. However, this trade-off is not proportional, as MAST achieves 69\% higher precision and 28\% higher F1 score across the full set of commits. 

The overall trends holds across the individual datasets, except for three cases (*-lib-common, *-datastream-dump-*-extractor, and *-service-dataexport-download) where the baseline achieves a higher F1 score. In all three cases, the difference comes down to a major difference in recall between the approaches. In 10 datasets, due to the difference in recall, the baseline also has a higher F2 score than MAST, as the F2 score places a high emphasis on recall. In some situations, recall may be of very high importance to a tester---i.e., they are willing to sift through many false positives to ensure they do not miss any true positives. Overall, MAST achieves the best balance of results, but it may miss some tests in such situations.

\begin{tcolorbox}[
    title=MAST versus Baseline on Positive Cases (RQ1),
    fonttitle=\bfseries,
    colback=gray!5,
    colframe=black!60,
    arc=2pt,
    boxrule=0.5pt,
    left=8pt, right=8pt, top=6pt, bottom=6pt,
    width=\linewidth
]
MAST achieves a much higher precision than the baseline, leading to better accuracy, F1, and F2 scores. However, there is a trade-off between precision and recall, with the baseline yielding higher recall than MAST. 
\end{tcolorbox}

\subsection{Comparison of MAST and Baseline in Negative Cases (RQ2)}

Table~\ref{tab:results_rq2_3} lists results aggregated across all projects. To address RQ2, we focus on negative commits---i.e., where a production code change did not lead to subsequent test maintenance. We also focus on the baseline (first row) and the full MAST framework (final row). In negative cases, there are only FP (tests predicted as needing maintenance, when no such maintenance is needed) and TN (tests correctly predicted as not needing maintenance). Therefore, we only list the number of FP and the resulting accuracy. For negative commits, we see that MAST was much more effective than the baseline. There was an 83\% reduction in false positives in such cases, leading to a 6\% improvement in accuracy.  

\begin{tcolorbox}[
    title=MAST versus Baseline on Negative Cases (RQ2),
    fonttitle=\bfseries,
    colback=gray!5,
    colframe=black!60,
    arc=2pt,
    boxrule=0.5pt,
    left=8pt, right=8pt, top=6pt, bottom=6pt,
    width=\linewidth
]
In cases where test maintenance was not required, MAST yields far fewer false positives than the baseline, resulting in a higher accuracy. 
\end{tcolorbox}

\begin{table*}[!t]
\centering
\caption{Comparison of LLM-based and a traditional heuristic-based fusion, aggregated across all projects.}
\label{tab:results_rq3_fusion}
\scriptsize
\begin{tabular}{lccccc}
\toprule
\textbf{Workflow} & \textbf{Precision} & \textbf{Recall} & \textbf{Accuracy} & \textbf{F1 Score} & \textbf{F2 Score} \\ 
\midrule
LLM Fusion w/o Post-Check & 0.355 $\pm$ 0.004 & \textbf{0.778 $\pm$ 0.018} & 0.854 $\pm$ 0.001 & 0.487 $\pm$ 0.007 & \textbf{0.628 $\pm$ 0.012} \\ 
Heuristic Fusion w/o Post-Check & \textbf{0.612 $\pm$ 0.002} & 0.500 $\pm$ 0.002 & \textbf{0.927 $\pm$ 0.000} & \textbf{0.550 $\pm$ 0.002} & 0.519 $\pm$ 0.002 \\ \midrule 
MAST (LLM Fusion) & 0.621 $\pm$ 0.003 & \textbf{0.613 $\pm$ 0.014} & 0.932 $\pm$ 0.001 & \textbf{0.617 $\pm$ 0.009} & \textbf{0.615 $\pm$ 0.012} \\ 
MAST (Heuristic Fusion) &  \textbf{0.736 $\pm$ 0.002} & 0.421 $\pm$ 0.005 & \textbf{0.935 $\pm$ 0.000} & 0.536 $\pm$ 0.004 & 0.460 $\pm$ 0.004 \\ 
\bottomrule
\end{tabular}
\end{table*}

\subsection{Ablation Study (RQ3)}

The goal of RQ3 is to understand how MAST's individual components and their interaction affect the performance of the full framework. To answer this question, we compare five workflows. First, we look at the three analyses of information sources (semantic analysis---i.e., the baseline---lexical analysis, and static analysis). Then, we examine MAST without applying the post-check analysis---i.e., we assess the fusion of the three information sources. Finally, we can compare to the final output of MAST (i.e., after applying the final post-check analysis). Table~\ref{tab:results_rq2_3} lists results for each workflow aggregated across all positive and negative commits. 

\subsubsection{Comparison of Information Sources}

Comparing the analyses of three information sources (semantic, lexical, and static analysis) on positive commits, we see that the static analysis yields a higher precision than the other two sources (0.502) and a balanced recall, leading to the highest F1 score of the three analyses. 
As noted previously, the semantic analysis yields a substantially higher recall---and somewhat higher F2 score---than the other two methods. However, on negative commits, it yields many more false positives than the other two information analyses. In negative cases, the lexical and static analysis yield similar performance, with a slight edge to the static analysis. 

While the lexical analysis yields lower performance than the other methods, when we examine the tests recommended by each information source qualitatively, there are some TP results suggested uniquely by each of the three analyses. Therefore, we see value in retaining each information source.

\subsubsection{Effect of Fusion}

From Table~\ref{tab:results_rq2_3}, in positive cases, we can see that the fusion---before the post-check---yields a lower precision than any of the individual analyses. However, it also has the highest recall (and F2 score) of any workflow, including the full MAST framework. Because we use an LLM to perform the fusion---rather than taking a simple union or intersection of the three sources---we achieve more nuanced results. 

The fusion yields a relatively inclusive list of test cases from the three sources, but still filters some cases with limited support. For example, the lexical analysis suggests a large number of FPs (hence its low precision) that were filtered by the fusion step. The fusion results in a high number of true positives---achieving high recall---but also many false positives that must be filtered by the post-check---i.e., the precision is relatively low. We see a similar effect in negative cases, where the fusion yields more false positives and---as a result---a lower accuracy than the individual information analyses because it produces a relatively inclusive list of recommendations from the individual analyses. 

\subsubsection{Effect of Post-Check}

In MAST, the post-check analysis takes each prediction from the fused list and compares the test code to the change in the production code to make one final assessment of the relevance of the suggestion. The intention of this analysis is to filter spurious suggestions from the fused list. From Table~\ref{tab:results_rq2_3}, and as discussed earlier, we can see that the post-check is successful in this task. In both positive and negative cases, the number of false positives is far lower for the full MAST framework (i.e., after post-check) than after the fusion, resulting in improved precision, accuracy, and F1 score. There is some cost to the recall as well, but less than the relative improvement in the precision. 

In cases where recall is highly prioritized, a tester may wish to inspect the fusion results instead of---or in comparison with---the full results of the MAST framework. However, given the number of false positives and the relative similarity in F2 score, the cost of inspecting all cases in the fused list must be considered. 

\subsubsection{LLM-Based Versus Traditional Fusion Methods}

A hypothesis underlying the design of MAST was that LLM-based fusion could yield superior results to traditional, heuristic-based fusion methods. We assessed this hypothesis by also implementing a version of MAST with a traditional fusion method, where a test is included in the merged list as long as it is in more than one of the fused lists. The comparison between the two fusion methods before the post-check, as well as the full MAST framework with each fusion method, is shown in Table~\ref{tab:results_rq3_fusion}. 

From the results, we can again see the tension between precision and recall. The heuristic fusion yields a higher precision, accuracy, and F1 score than the LLM-based fusion. However, the LLM-based fusion yields a similarly substantial increase in recall and F2 score over the heuristic method. This further shows that the LLM-based fusion is not a simple merging of the individual lists, but the product of a more complex analysis. 

We can see the full effect when each fusion is integrated into the full MAST framework, with its subsequent post-check. MAST with the LLM-based fusion yields a substantially higher recall, F1, and F2 score, as well as a similar accuracy---while MAST with the heuristic fusion yields a higher precision. Overall, while neither method yields better results in all measurements, the LLM-based fusion ultimately leads to a better overall performance due to its recall---i.e., its high true positive rate. 

\begin{tcolorbox}[
    title=Ablation Study (RQ3),
    fonttitle=\bfseries,
    colback=gray!5,
    colframe=black!60,
    arc=2pt,
    boxrule=0.5pt,
    left=8pt, right=8pt, top=6pt, bottom=6pt,
    width=\linewidth
]
Each information source yields different, but potentially valuable, maintenance suggestions. The fusion agent effectively merges the sources, and the post-check prunes many false positives from the merged list (albeit at some cost in recall). The superior results of MAST in precision, accuracy, and F1 score over the individual workflows show that each agent contributes to the overall effectiveness of MAST. However, in situations where recall is highly prioritized, a tester may also wish to inspect the fusion results prior to the post-check. 
\end{tcolorbox}

\section{Discussion}\label{sec:disc}
\subsection{Implications of the Results}

From the results, we can see that MAST achieves a higher precision, accuracy, F1, and F2 score than the baseline on positive commits and a higher accuracy on negative commits---due to substantially fewer false positives. Like in many machine learning tasks, there is a natural tension between precision and recall. The baseline---as well as the fusion results before applying the post-check---offer a higher recall than the full MAST framework. However, this trade-off is not symmetrical. MAST achieves a substantially higher precision with only some loss in recall. Overall, therefore, we recommend the use of MAST over the baseline or the fusion results except in cases where any loss in recall is unacceptable.

Our findings highlight the complementary nature of the different analysis techniques. The semantic analysis captures the intentions underlying each test case, but also suffers from lower precision by operating on natural language interpretations of the test code. The fusion overcomes this limitation by also incorporating the lexical analysis---which captures syntactic similarity of the code---and the static analysis---which captures structural dependencies between tests and production code. The post-check then successfully filters cases from the fused list of test cases that lack sufficient support for their inclusion. MAST effectively fuses information from different sources, offering evidence that LLM-based analysis fusion could assist in other software engineering tasks as well. 

\subsection{Recommendations for Practitioners}

A natural question is---are the results of MAST good enough to use the framework in practice? 
To contextualize the current results of MAST, an average of 6.5 test cases are updated in each positive commit in the ground truth. When MAST is executed for a commit, it will generally yield four correct recommendations (TP), two incorrect recommendations (FP), and miss two tests that should have been identified (FN). 

These results are promising. About two-thirds of MAST's recommendations are correct, and the number of false positives is not burdensome. The main barrier to relying on MAST are the false negatives---the missed cases. If MAST were used as the sole decision-maker for test maintenance, then the project quality could still suffer. These results suggest that there is a strong potential for applying LLM-based tools in industrial test maintenance, but also as an indication that there is still room for improvement.

Our recommendation is to use MAST as a way to accelerate the test maintenance process, but not as the sole determination of what tests to update and not as a total replacement for human judgment. We envision two primary use-cases for MAST:
\begin{itemize}
    \item MAST could be manually invoked after the production code is updated, but before it is committed to the repository, by the developer. The developer could use its recommendations, along with their own intuition or other tools, to perform any required test maintenance.
    \item MAST could be invoked in the continuous integration pipeline after the production changes are committed (along with any planned test maintenance) as a way to identify potentially missed maintenance cases. Its recommendations could then be offered as advice to the developer. 
\end{itemize}
In both cases, MAST can ease the burden of test localization, but its predictions should be taken as recommendations that are validated by the developer.

\subsection{Threats to Validity}

\subsubsection{External Validity}

We conducted our evaluation at one company, Ericsson AB, and focused on one programming language, Java. Therefore, some aspects of this study may not generalize beyond this specific context. MAST is not closely coupled to any particular technology or project at Ericsson, nor to specific aspects of the Java language (except for some parsing tools). However, the specific performance trends may not hold on all projects or all languages. We partially mitigated this threat by curating commits from 21 different projects, covering diverse domains and project sizes. Therefore, because the projects used in our evaluation were of a similar scope, we speculate that our results will hold---at least---on other small-to-medium sized Java projects. In the future, we will also consider open-source projects and additional languages.

\subsubsection{Internal Validity}

We made two assumptions when collecting commits. First, we assumed that tests were updated because of the production code changes in the same commit (i.e., code and tests co-evolved). Second, we assumed that no tests were missed by developers. Both assumptions are in-line with previous research, but may also be incorrect at times. Some tests may have been missed or were not updated until later commits. This means that some false positives may actually be valid suggestions---something we witnessed in our past work~\cite{liu2025exploringintegrationlargelanguage}. The performance of MAST in real-world situations may differ somewhat from in the evaluation. However, in some cases, ``false positives'' may still offer valuable suggestions to developers.

\subsubsection{Construct Validity}

We evaluated MAST against a baseline drawn from our prior work~\cite{liu2025exploringintegrationlargelanguage}, but we did not include the approaches proposed by Hu et al.~\cite{hu2023identify} or Chi et al.~\cite{chi2025reaccept}. However, their approaches cannot be executed under equivalent conditions (the limitations of prior work were discussed in Sections~\ref{sec:intro}--\ref{sec:background}), and the effort needed to re-implement their approaches for our context could not be justified. 

\section{Conclusion}\label{sec:conclusion}
In this study, we presented MAST, a multi-agent framework that performs test localization following a change to the production code. MAST advances the state-of-the-art by integrating multiple code analysis techniques---including static, lexical, and semantic analysis---and through our focus on a realistic use and evaluation setting---i.e., MAST operates on standard input formats, is not limited to single-method changes, does not require an explicit mapping of production and test code, and has been evaluated on cases where test maintenance was and was not required in the ground truth.

MAST was more effective than a baseline from our prior work, achieving higher precision, accuracy, F1, and F2 scores. We witnessed a trade-off between precision and recall. However, MAST still yields a far higher precision than the baseline, with only some loss in recall. Our ablation study demonstrates the value of each integrated analysis in producing the final set of recommendations.

In future work, we plan to improve the performance of MAST through techniques such as prompt optimization and the integration of long-term memory, expand our analysis to open-source repositories, explore languages other than Java, and expand MAST to perform maintenance actions on the localized test cases.

\smallskip\noindent\textbf{Acknowledgments:} Support for this research was provided by the Wallenberg AI, Autonomous Systems and Software Program (WASP) funded by the Knut and Alice Wallenberg Foundation.

\section{Data Availability Statement}\label{sec:availability}

The current version of MAST is available at \url{https://github.com/Roy19921010/MAST}. Because this framework may evolve in the future, we also make available a supplementary package at \url{https://doi.org/10.5281/zenodo.19680703} containing a snapshot of the version of MAST used in this study, as well as the experiment results for all datasets. The Ericsson repositories used in our study cannot be made availability due to confidentiality constraints. 

\bibliographystyle{ACM-Reference-Format}
\bibliography{main}

\end{document}